\documentclass[superscriptaddress, 
nobibnotes, reprint,floatfix, pra, 
]{revtex4-2}
\usepackage{upgreek}
\usepackage{amsmath,braket}
\usepackage{graphicx}
\usepackage{mathtools}
\usepackage{color}
\usepackage{xcolor}
\usepackage{bm} 
\usepackage{setspace}
\newcommand{\RNum}[1]{\uppercase\expandafter{\romannumeral #1\relax}}
\usepackage{comment}
\usepackage{soul}
\usepackage{ulem}

\usepackage[mathlines]{lineno}
\begin{document}


\title{Two-photon interference between mutually-detuned resonance fluorescence signals scattered off a semiconductor quantum dot}

\author{Guoqi~Huang\textsuperscript{\#}}
\affiliation{Beijing Academy of Quantum Information Sciences, Beijing 100193, China}
\affiliation{School of Science, Beijing University of Posts and Telecommunications, Beijing 100876, China}
\author{Jian Wang\textsuperscript{\#}}
\affiliation{Beijing Academy of Quantum Information Sciences, Beijing 100193, China}
\author{Ziqi Zeng}
\affiliation{Beijing Academy of Quantum Information Sciences, Beijing 100193, China}
\affiliation{Beijing National Laboratory for Condensed Matter Physics, Institute of Physics, Chinese Academy of Sciences, Beijing 100190, China}
\affiliation{University of Chinese Academy of Sciences, Beijing 101148, China}
\author{Hanqing~Liu}
\affiliation{State Key Laboratory of Optoelectronic Materials and Devices, Institute of Semiconductors, Chinese Academy of Sciences, Beijing 100083, China}
\affiliation{Center of Materials Science and Optoelectronics Engineering, University of Chinese Academy of Sciences, Beijing 100049, China}
\author{Li~Liu}
\author{Weijie~Ji}
\author{Bang~Wu}
\affiliation{Beijing Academy of Quantum Information Sciences, Beijing 100193, China}
\author{Haiqiao~Ni}
\author{Zhichuan~Niu}
\email{niuzc@semi.ac.cn}
\affiliation{State Key Laboratory of Optoelectronic Materials and Devices, Institute of Semiconductors, Chinese Academy of Sciences, Beijing 100083, China}
\affiliation{Center of Materials Science and Optoelectronics Engineering, University of Chinese Academy of Sciences, Beijing 100049, China}
\author{Rongzhen Jiao}
\affiliation{School of Science, Beijing University of Posts and Telecommunications, Beijing 100876, China}
\author{Davide G. Marangon}
\affiliation{Department of Engineering of Information, Padua University, Italy}
\author{Zhiliang~Yuan}
\email{yuanzl@baqis.ac.cn}
\affiliation{Beijing Academy of Quantum Information Sciences, Beijing 100193, China}

\date{\today}

\begin{abstract}
The radiative linewidth of a two-level emitter (TLE) fundamentally limits the bandwidth available for quantum information processing.  
Despite its importance, no prior experiment has systematically examined how driving detuning affects the indistinguishability of photons scattered from a TLE—a parameter critical for photonic quantum computing.  
Here, we perform post-selective two-photon interference measurements between mutually detuned resonance fluorescence signals from an InAs quantum dot embedded in a micropillar cavity.  
At small mutual laser detunings ($\leq 0.5$~GHz), the results are accurately described by the pure-state model [\textit{Nat. Commun.} \textbf{16}, 6453 (2025)], which treats all resonance-fluorescence photons as spontaneous emission.  
At larger detunings, we uncover an anomalous feature in the two-photon interference, where the normalised second-order correlation function under orthogonal polarisations yields $g^{(2)}_{\perp}(0) < 0.5$.
\end{abstract}

\maketitle

Impinging upon a semi-transparent beam-splitter, two indistinguishable single photons will coalesce into either of the two output ports, and never exit separately. Referred to as Hong-Ou-Mandel (HOM) effect~\cite{Hong1987}, this phenomenon is the foundation for quantum information protocols such as repeater-assisted quantum communication~\cite{Azuma_2023} and linear-optic quantum computing~\cite{Kok_2007}.
Single photons can be generated through exciting an atom-like system~\cite{Grangier_1986, michler_quantum_2000,yuan_2002}, which by nature emits one photon at a time. In this regard, semiconductor quantum dots (QDs) are arguably the most attractive platform because they can be integrated into monolithic structures for cavity enhancement~\cite{michler_quantum_2000,pelton_efficient_2002,liu_high_2018,tomm2021bright,ding2025high} and highly indistinguishable single photons~\cite{liu_solid-state_2019,zhaiQuantumInterferenceIdentical2022}.

Resonant excitation into the lowest excited state allows a QD system to be simplified as a two-level emitter (TLE)~\cite{muller2007resonance}.
Under a weak coherent pump (Heitler regime), the resulting resonance fluorescence (RF) signal locks to the driving laser's coherence~\cite{nguyen2011ultra,matthiesen2013phase,Wells2023, wang2023} while exhibiting single-photon characteristics.
To a first approximation, the QD appears to behave like a passive and elastic scatterer,  which led to an optimistic idea to circumvent environment-induced dephasing inherent in all solid-state systems and to generate single photons with indistinguishability ``fundamentally" defined by the excitation laser~\cite{matthiesen2013phase}.
A question then arises. In the presence of an excitation detuning, will the single photons remain indistinguishable from their counterparts generated under strictly resonant excitation? An affirmative answer would be favorable for quantum information processing because it will then permit arbitrary modulation of the driving laser and the shaping of the RF beam accordingly without loss of photon indistinguishability.  However, no experiment has so far been dedicated to explicitly examining this question, while substantial efforts have been made to verify photon indistinguishability between remote quantum emitters~\cite{patelTwophotonInterferenceEmission2010,flagg_2010,lettow_2010,giesz2015cavity,weber2019two}.

\begin{figure*}[t]
\centering
\includegraphics[width=1.8\columnwidth]{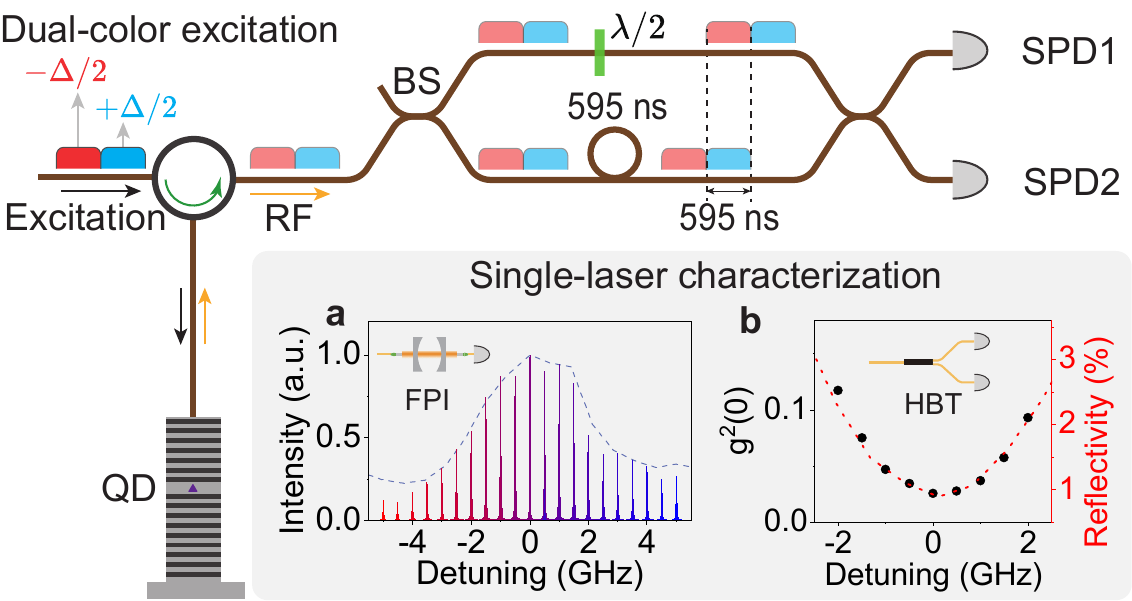}	
\caption{\textbf{Dual-color two-photon interference (TPI) setup.} The RF signals are collected in the same polarization as the excitation lasers, and no spectral filtering is applied to them in any of the TPI experiments.
\textbf{a}, High-resolution RF spectra measured with a Fabry-P\`erot interferometer (FPI);  The dashed line represents the QD-micropillar reflectance spectrum measured under $\bar{n} = 0.05$.
\textbf{b}, The auto-correlation function $g^{(2)}(0)$ values measured under $\bar{n} = 0.05$ using a Hanbury Brown-Twiss setup (HBT). The red dashed line shows the cavity reflectivity under a strong excitation of 465~nW, corresponding to $\bar{n} = 139$. BS: beam splitter; SPD: single-photon detector.}
	\label{fig:setup}
\end{figure*}

\begin{figure*}[t]		
\centering
\includegraphics[width=1.8\columnwidth]{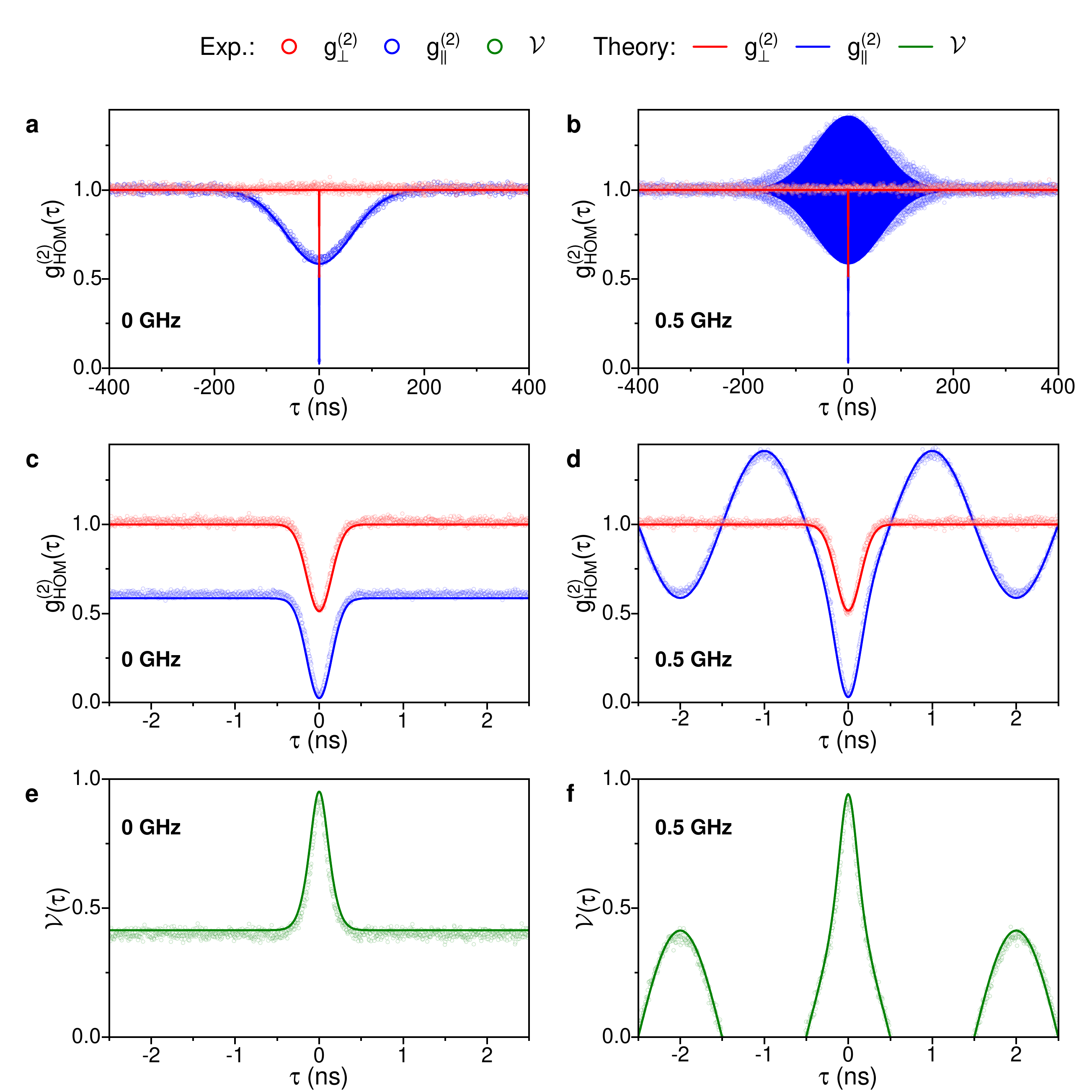}	
\caption{\textbf{Dual-color HOM interference results under two different mutual detunings: $\Delta =0$~GHz (left column) and $\Delta =0.5$~GHz (right column).}
\textbf{a}, \textbf{b}, HOM interference traces for the parallel (blue) and orthogonal (red) polarization settings;  \textbf{c}, \textbf{d}, Magnified view of the data shown in panels \textbf{a} and \textbf{b};
\textbf{e}, \textbf{f}, TPI visibility. In all panels, experimental data are shown as dots while theoretical simulations are displayed as solid lines.  In this set of measurements, the excitation flux is fixed at $\bar{n} = 0.05$.}
	\label{HOM}
\end{figure*}

Here, we subject an InAs QD-micropillar device to dual-color excitation, i.e., from two different wavelengths, using the setup shown in Fig.~\ref{fig:setup},
and evaluate indistinguishability of mutually-detuned RF signals using post-selective two-photon interference (TPI)~\cite{patelPostselectiveTwoPhotonInterference2008}.
Two continuous-wave (CW) lasers (not shown) are each modulated into square pulses of 595~ns duration, separated by 595~ns off periods, and are then temporally interleaved to sequentially excite the QD device. Symmetrical detuning is applied:  one laser is detuned by $+\frac{\Delta}{2}$ while the other by $-\frac{ \Delta}{2}$ from the QD resonance ($\nu_0$). The resulting RF tracks the lasers' frequencies and consists of RF signal pulses of alternating colors. 
Without applying any spectral filtering, the RF signals enter an asymmetric Mach-Zehnder interferometer (AMZI) of 595~ns delay that matches the laser carving period, so that the RF signals of different colors are temporally aligned when meeting at the AMZI's exit beam splitter. The interference results are registered by a pair of superconducting nanowire single photon detectors (SPDs) of 59~ps temporal resolution.

The InAs QD sample was fabricated from a wafer grown by molecular beam epitaxy on a (001) semi-insulating GaAs substrate and embedded into a double distributed-Bragg-reflector (DBR) micropillar cavity with a diameter of 2.4~$\mu$m.
A confocal microscope (not shown) collects the RF at the same polarization as the excitation laser, and no spectral filtering is applied in the RF signal path, thanks to the low cavity reflection of the device~\cite{wu2023,wang2023}. The QD transition we study corresponds to a neutral exciton state, which exhibits a small fine structure splitting ($\mathrm{FSS} \approx 0.91$ GHz), see Fig.~\ref{fig:fig10}. The device operates in the weak-coupling regime with a cooperativity $C = \frac{g^2}{\kappa \gamma} = \frac{F_p}{2} \approx 5.45$, corresponding to a Purcell factor of approximately 10.9. Here, the coupling strength between the cavity and the QD exciton state is $g\approx 6.2$~GHz, and the cavity damping rate is $\kappa \approx 35$~GHz (corresponding to a quality factor of 9350). The sample is cooled by a closed-cycle cryostat to 5.6~K to ensure the QD's resonance with the micropillar cavity of 911.485~nm. Under pulsed resonant excitation, we characterize the QD to have a radiative lifetime ($T_1$) of $74$~ps, corresponding to a natural linewidth of $\gamma_\parallel := \frac{1}{2\pi T_1} \approx 2.15$ GHz. In all subsequent experiments, we denote the excitation flux upon the device surface using the relation $\bar{n} = P_\mathrm{in}T_1/h\nu$, where $P_\mathrm{in}$ corresponds to the CW incident optical power. 

Further experimental details and measurement data, including the pulse-carving of the excitation laser as well as the determinations of the QD's radiative lifetime ($T_1$) and fine-structure splitting, are provided in Appendix~B.

Before the dual-color experiment, we characterized the QD under excitation by a single laser. The excitation flux is kept at $\bar{n} = 0.05$, which is conservatively set low to operate in the Heitler regime.
Fig.~\ref{fig:setup}\textbf{a} shows a set of high-resolution RF spectra measured by a Fabry-P\`erot interferometer (FPI).
At each detuning,  the RF peak inherits the laser's optical frequency and has a full-width at half-maximum width (FWHM) that is limited by the FPI resolution ($20$~MHz).
The peak intensity decreases as the detuning widens, which is a result of the reduced cavity enhancement. At the measurement step-size of 0.5~GHz, the adjacent RF spectra are fully distinguished.
Fig.~\ref{fig:setup}\textbf{b} shows the corresponding auto-correlation function results, measured with a Hanbury Brown-Twiss (HBT) setup. 
Across the detuning range, $g^{(2)}(0)$ is consistently lower than 0.5,  indicating the non-classical nature of each RF signal. The lowest  $g^{(2)}(0) = 0.026$ is obtained at 0-detuning. It gradually deteriorates as the detuning increases, and exceeds $0.10$ when the detuning reaches $\pm 2$~GHz. We attribute the deterioration in single-photon purity to the increased laser background, as evidenced by the cavity reflectivity (red dashed line), which was measured under strong laser excitation.%

We use the setup in Fig.~\ref{fig:setup} for dual-color excitation experiments.
By rotating the half-wave plate ($\lambda/2$) in the long arm of the AMZI, the polarization alignment between the incoming photon beams can be made either parallel or orthogonal.  In the latter case, photons are fully distinguishable.
We measure the normalized second-order correlation functions under both polarization settings,  from which we calculate the TPI visibility ($\mathcal{V}$) using~\cite{patelPostselectiveTwoPhotonInterference2008}
\begin{equation}
    \mathcal{V(\tau)} = \frac{g^{(2)}_{\perp}(\tau) - g^{(2)}_{\parallel}(\tau)}{g^{(2)}_{\perp}(\tau)}, \label{eq:visibility}
\end{equation}
where $\tau$ represents the coincidence time delay while $g^{(2)}_\perp(\tau)$ and $g^{(2)}_\parallel(\tau)$ are the normalized second-order correlation functions under the orthogonal and parallel polarization settings, respectively.

As a control experiment, we set the two lasers to be nominally identical in wavelength, i.e., $\Delta = 0$ GHz, and  Fig.~\ref{HOM}\textbf{a} shows the measured HOM traces (dots). Here, the coincidence probabilities are normalized to the average rate at long delays ($\tau > 200$~ns).
With orthogonal polarizations, the photons are fully distinguishable and therefore no interference takes place. Hence, the corresponding $g^{(2)}_\perp (\tau)$ trace displays a single, sharp dip with a depth of $\sim$0.5 and a width of about 100~ps, arising from the single-photon nature of the RF streams. When the polarizations are aligned, photons become largely indistinguishable, two-photon coalescence takes place thus resulting in a more pronounced HOM dip at $\tau = 0$. Additionally, the correlation function exhibits a broad suppression extending over a timescale of $\pm 200$~ns, corresponding to the mutual coherence time of the two pump lasers. Over nanosecond timescale (Fig.~\ref{HOM}\textbf{c}), $g_\perp^{(2)}$ and $g_\parallel ^{(2)}$ exhibit  flat values for $\tau \gg T_1$ but with $g_\parallel ^{(2)}$ being considerably lower than $g_\perp ^{(2)}$. This is a common observation in the HOM measurements for resonantly excited QDs~\cite{Wells2023}. Using Eq.~1, we obtain the TPI visibility shown in Fig.~\ref{HOM}\textbf{e}.

Next, we measure the second-order HOM correlation functions under dual-color excitation, with the result  of a $\Delta = 0.5$~GHz detuning shown in
Fig.~\ref{HOM}\textbf{b}.  While the $g^{(2)}_\perp$ data is qualitatively identical to the non-detuned case, the $g^{(2)}_\parallel$ data exhibits symmetrical oscillations around $g^{(2)}_{HOM} = 1$. The oscillation amplitude decreases as the delay enlarges, and its lower envelope closely resembles the non-detuned trace.
Under a magnified view (Fig.~\ref{HOM}\textbf{d}),  the oscillation is discerned to have a period of $1/\Delta$, similarly to the asynchronous TPI result between frequency-offset laser fields~\cite{zhou_experimental_2023}.
As a display of the HOM effect, photon coalescence leads $g^{(2)}_\parallel (0)$ to be substantially lower than the oscillation minima.
Figure~\ref{HOM}\textbf{f} shows the TPI visibility, whose trace has a substantially different appearance from that in Fig.~\ref{HOM}\textbf{e}.
The $\mathcal{V}$ value oscillates (negative data not shown) with the oscillation maxima upper-bounded to the TPI baseline measured in the non-detuning case. Nevertheless, the $\mathcal{V}$ maximum at $\tau = 0$ is about identical to that of the non-detuned case.

To explain the above observations, we resort to the recently introduced RF model~\cite{wang2023} that uniquely treats all the RF photons as a result of absorption and re-emission. Considering the TLE and its photon emission together,  this RF model adopts a pure-state description for the joint system of the TLE and its emission, although the TLE or the RF emission \textit{separately} is well known to be in a mixed state.
As derived in Appendix~A1, the quantum state of the light-matter system can be written as
\begin{equation}
 \ket{\psi} 
 = \sqrt{p_0}\ket{0,g} + \sqrt{p_1} {\color{black}e ^ {-i(2\pi \nu t + \theta)}} \frac{\ket{0,e} + \ket{1,g}}
 {\sqrt{2}}, 
  \label{eq:photonstate_1}
\end{equation}
where the state $\ket{0}$ ($\ket{1}$) corresponds to the absence (presence) of a \textit{freshly} emitted photon,  $\ket{g}$ ($\ket{e}$) the TLE's ground (excited) state and $\nu$ the driving laser's frequency, which can be detuned from the TLE's transition frequency $\nu_0$. The parameter $\theta$ represents the detuning-dependent phase shift, which, however, produces no observable effect in the measurements presented in this work. The terms $p_0$ and $p_1$ ($p_0+p_1=1$) represent the probabilities of the TLE system occupying the system ground state $\ket{0,g}$  and the joint excited state $\left(\ket{0,e} + \ket{1,g}\right)/\sqrt{2}$, respectively.  The latter is in a superposition form conveying the insight that at any instant of time, it is not possible to know whether the TLE stays in its excited state $\ket{e}$ or has decayed to the ground state $\ket{g}$ and emitted a photon~\cite{wang2023}.  Naturally, $\ket{1}$ represents a photon emitted by the TLE, implying that the property of the photons is independent of excitation detuning. 

Equation~\ref{eq:photonstate_1} provides a unified description of resonance fluorescence, reproducing both the laser-like spectrum and photon antibunching shown in Fig.~\ref{fig:setup}\textbf{b} and~\textbf{c}. First, the quantum state can be shown to exhibit photon antibunching, as characterized by the second-order correlation function 
\[
g^{(2)}(\tau)=\frac{\langle a^{\dagger}(t)a^{\dagger}(t+\tau)a(t+\tau)a(t)\rangle}
{\langle a^{\dagger}(t)a(t)\rangle^2},
\]
which yields $g^{(2)}(0)=0$ owing to the absence of two-photon probability.  Second, 
we can infer the power spectrum by working out the first-order coherence function,
\[
g^{(1)}(\tau)=\frac{\langle a^{\dagger}(t)a(t+\tau)\rangle}{\langle a^{\dagger}(t)a(t)\rangle}
= p_0 e^{-i2\pi\nu\tau},
\]
valid for $T_1 \ll \tau \ll T_L$, where $T_L$ denotes the driving laser's coherence time.  
The Fourier transform of $g^{(1)}(\tau)$ then yields a power spectrum containing a narrow, laser-like component centered at $\nu$ with a spectral weight of $p_0$. Specifically, $p_0$ is dependent on the excitation power
\begin{equation}
    p_0 = \frac{1}{1 + 2\eta_{ab}\bar{n}}, 
\label{eq:p0}
\end{equation}
where $\eta_{ab}$ denotes the TLE’s absorption efficiency in the weak-drive regime.

\begin{figure}[tb]		
\centering
\includegraphics[width=1\columnwidth]{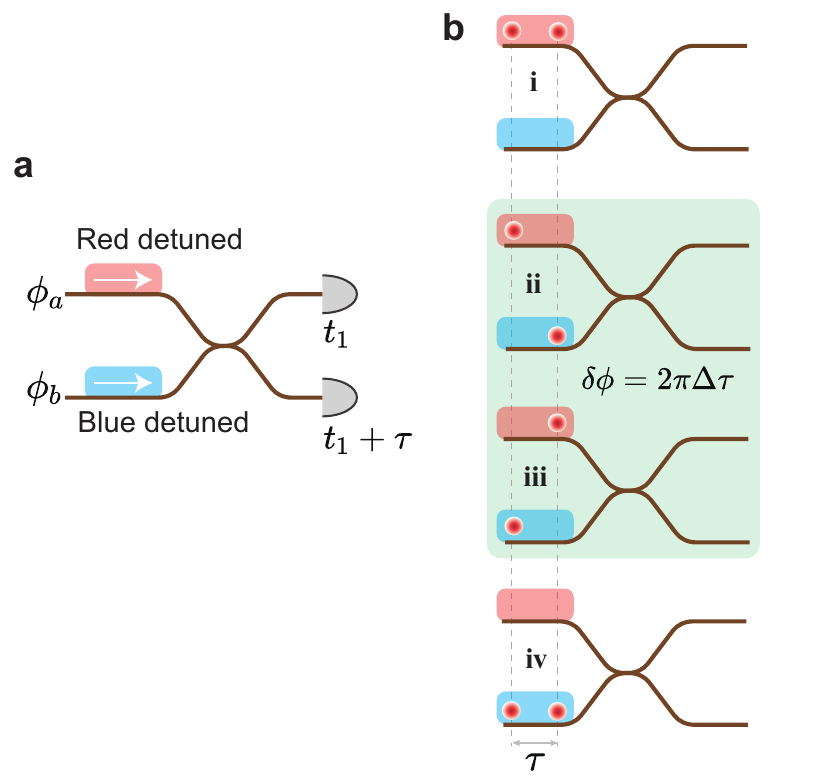}	
\caption{\textbf{Visualization of two-photon coincidence contributions.} \textbf{a}, Two mutually-detuned RF beams, with global phases of $\phi_a$ and $\phi_b$, respectively,  meet at a 50/50 beam splitter, and their interference is recorded by two single-photon detectors; \textbf{b}, Four possible cases involving two input photons may contribute to a coincidence event with the two detectors clicking respectively at time $t_1$ and $t_2 = t_1 + \tau$. Cases (\textbf{ii}) and (\textbf{iii}), in which each input arm contributes one photon, share the common global phase ($\phi_a + \phi_b$) and therefore their quantum probability amplitudes interfere. For simplicity, cases involving three or more input photons are omitted from the diagram, but they are fully accounted for in Eq.~\ref{eq:HOM_all}.}
\label{fig:TPI-Schematic}
\end{figure}

The dual-color HOM data can be elegantly explained through the schematic shown in Fig.~\ref{fig:TPI-Schematic}. Without loss of generality, we fix the entry of the red (blue) detuned RF signal to the top (bottom) input of the 50/50 beam splitter. Their global phases are respectively depicted as $\phi_a$ and $\phi_b$, and are not mutually locked.
A coincidence of photon clicks at $t_1$ and $t_2 = t_1 + \tau$ by two detectors can arise from four possible cases: both photons came from the same entry (\textbf{i} and \textbf{iv}) or they entered from different entries (\textbf{ii} and \textbf{iii}). 
Here, we temporarily ignore cases involving three or more input photons to avoid complications. Among the four possible cases, cases (\textbf{ii}) and (\textbf{iii}) share the same global phase ($\phi_a + \phi_b$) and therefore interference occurs after their transformation through the beam splitter. The phase difference is $2\pi \Delta \tau + \pi$, where $2\pi \Delta \tau$ arises from the RF detuning and an extra $\pi$ phase gains from the beam splitter.  
The interference will cause the total coincidence probability to oscillate around an intermediate value that is defined by cases (\textbf{i}) and  (\textbf{iv}).

In the single-photon limit that each input arm contributes at most one photon between time bins $t$ and $t + \tau$, i.e., $g^{(2)}(\tau) \to 0$, the coincidence probability amplitudes from cases (\textbf{ii}) and (\textbf{iii}) can cancel completely when $\Delta \cdot \tau$ is an integer (due to a $\pi$ phase shift from the beam splitter), but interfere constructively when $\Delta \cdot \tau$ becomes an half integer.
The HOM correlation function is phase-dependent:
\begin{equation}
g^{(2)}_{HOM}(\tau) = \frac{1-\mathcal{M}\cos {2\pi \Delta \tau}}{2}, \label{eq:Standard_HOM}
\end{equation}
where $\mathcal{M} = |\langle 1_{+} | 1_{-} \rangle|^2$ quantifies the indistinguishability between single-photon states $\ket{1_\pm}$ that are temporally-aligned but generated under opposite detunings. We remark that states $\ket{1_\pm}$ represent spontaneous emission and therefore their indistinguishability is affected by the QD environment fluctuation and/or phonon scattering~\cite{Iles-Smith2017}  but independent of the excitation detuning.
$\mathcal{M}$ can also be adjusted through the half-wave plate in the experimental setup (Fig.~\ref{fig:setup}).  With orthogonal polarizations, $\mathcal{M} = 0$ and $g^{(2)}_\perp (0)$ has a theoretical minimum of 0.5.

When the coincidence delay is much greater than the QD's exciton radiative lifetime, we arrive at the other limit that  $g^{(2)}(\tau) = 1$ and the probability of time bin $t + \tau$ containing a photon is independent of that at time bin $t$.
Notably, by using the pure-state of Eq.~\ref{eq:photonstate_1},  we can rigorously derive the second-order HOM correlation function for $\tau \gg T_1$ as
\begin{equation}
      g_{HOM}^{(2)}(\tau)
      =  \frac{1}{2} + \frac{1}{2} \left (1 - \mathcal{M} p_0^2 \cos {2\pi \Delta \tau} \right ). \label{eq:HOM_Async}
\end{equation}
The first term $\frac{1}{2}$ arises from cases (\textbf{i}) and (\textbf{iv}), i.e., both photons within a coincidence event came from the same beam splitter input.  The second term represents the interference between the quantum probability amplitudes of cases (\textbf{ii}) and (\textbf{iii}), but which can no longer cancel \textit{completely} because of the non-negligible probability of each input arm contributing two photons.
The detailed derivation of Eq.~\ref{eq:HOM_Async} is presented in Appendix~A2.

To model the situation of an arbitrary coincidence delay,  we bridge the above two limits using the auto-correlation function $g^{(2)}(\tau)$ and obtain
\begin{equation}
 \begin{aligned}
     g_{HOM}^{(2)}(\tau) =  & g^{(2)}(\tau) \times \left ( 1 - \frac{1}{2}\mathcal{M} p_0^2 
     \cos {2\pi \Delta \tau}  \right ) \\
   &  + (1 - g^{(2)}(\tau)) \times  \frac{1 - \mathcal{M} \cos{2\pi \Delta \tau}} {2}.
    \label{eq:HOM_all}
 \end{aligned}
\end{equation}
As check, one has that for  $\tau \gg T_1$, Eq.~\ref{eq:HOM_all} reduces to the form of Eq.~\ref{eq:HOM_Async} using $g^{(2)}(\tau) = 1$. At the other end for $\tau = 0$, ideal single-photon purity of $g^{(2)}(0) = 0$ leads the equation to standard post-selective HOM interference, i.e., Eq.~\ref{eq:Standard_HOM}.

We simulate the HOM experiments using  Eq.~\ref{eq:HOM_all}, with results shown as solid lines
in Fig.~\ref{HOM}. The auto-correlation function is directly taken from the experimental HBT data using $g^{(2)}(\tau) = \sqrt{g_+^{(2)}(\tau)g_-^{(2)}(\tau)}$, where $g_{\pm}^{(2)}(\tau)$ are the measured auto-correlation functions under the corresponding positive ($+\frac{\Delta}{2}$) or negative ($-\frac{\Delta}{2}$) excitation detuning.
We use $\mathcal{M}= 0.94$ ($\mathcal{M}= 0$) for the parallel (orthogonal) polarization setting.
 We also take into account the finite coherence time between the driving lasers, which deteriorates the TPI visibility at large delays. As reported in Figs.~\ref{HOM}\textbf{a}--\textbf{d}, the simulation faithfully reproduces all the measurement data for both polarization settings,  both excitation detunings and all delays.

\begin{figure}		
\centering
\includegraphics[width=1\columnwidth]{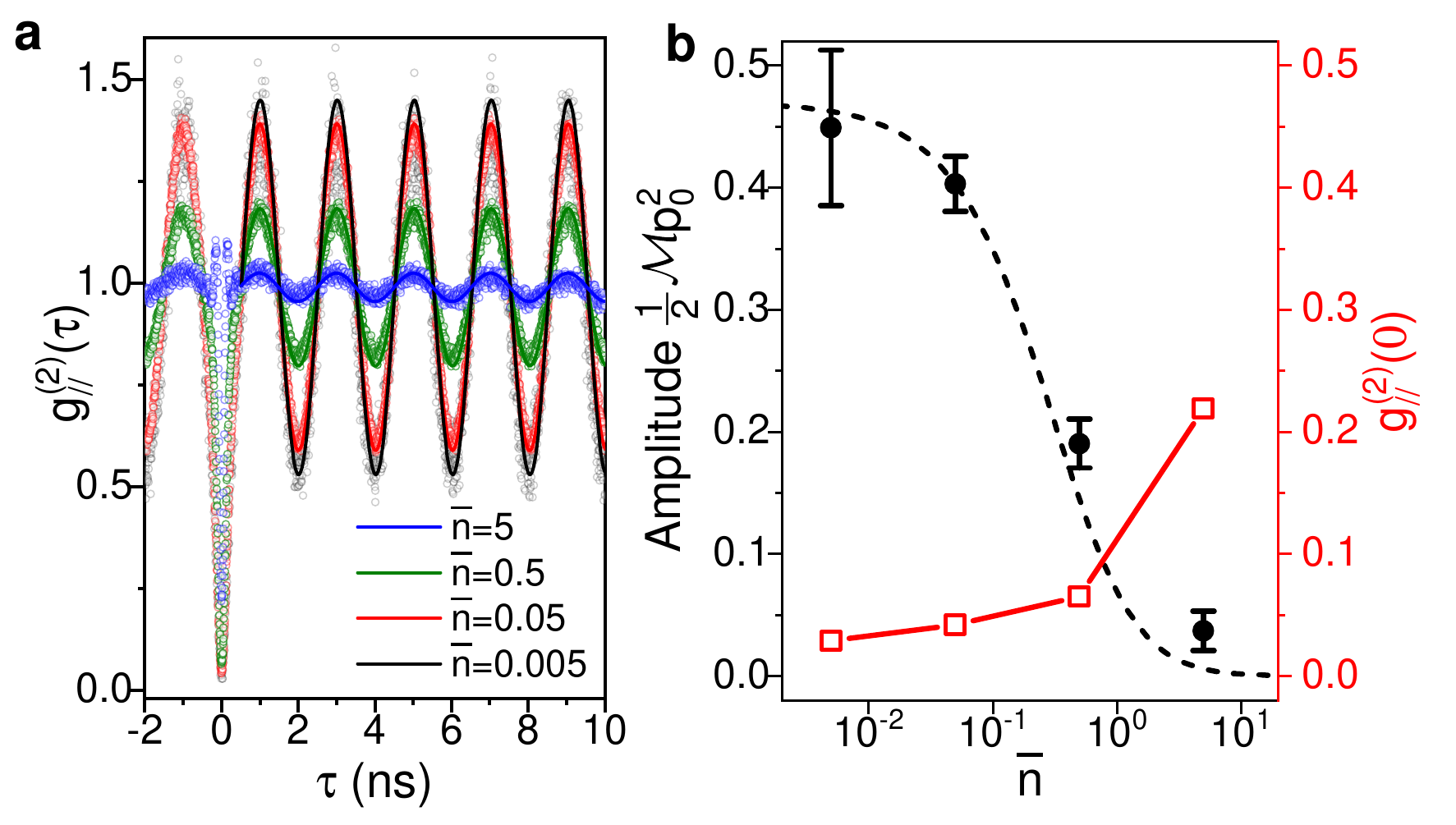}	
\caption{\textbf{Excitation flux dependence of the HOM correlation under the parallel polarization setting.}  The excitation detuning is fixed at $\Delta = 0.5$~GHz. \textbf{a}, HOM correlation traces. Symbols (solid lines) denote experimental data  (theoretical fits).
\textbf{b}, The oscillation amplitudes (solid dots) and the zero-delay HOM dip values (open squares), extracted from the data shown in panel \textbf{a}, along with the fit (dashed line) to the former using Eq.~\ref{eq:p0}.
}
	\label{power_dep}
\end{figure}

In the above simulation, we choose $p_0 = 0.920\pm 0.001$ corresponding to an excitation flux of $\bar{n} = 0.050\pm0.002$. As $p_0$ decreases with a strengthening driving intensity, we expect the oscillation amplitude of the detuned HOM correlation to dampen according to Eq.~\ref{eq:HOM_Async}. To test, we measure power-dependent dual-color HOM traces with the results (symbols) shown in Fig.~\ref{power_dep}\textbf{a}.  Fitting the correlation oscillations using Eq.~\ref{eq:HOM_Async} (solid lines), we extract the oscillation amplitude ($\frac{1}{2} \mathcal{M} p_0^2$) and plot its dependence on the excitation flux in Fig.~\ref{power_dep}\textbf{b}.  The measured dependence (symbol) is in agreement with the theoretical simulation (dashed line) using Eq.~\ref{eq:p0}. 
Figure~\ref{power_dep}\textbf{b} also plots the HOM dip, \( g^{(2)}_\parallel(0) \), which degrades monotonically with increasing excitation power but at a slower rate than the oscillation amplitude. The degradation arises from the single-photon purity, deteriorated by the increasing laser background at high pump power~\cite{wang2023}.

\begin{figure}[t]		
\centering
\includegraphics[width=.9\columnwidth]{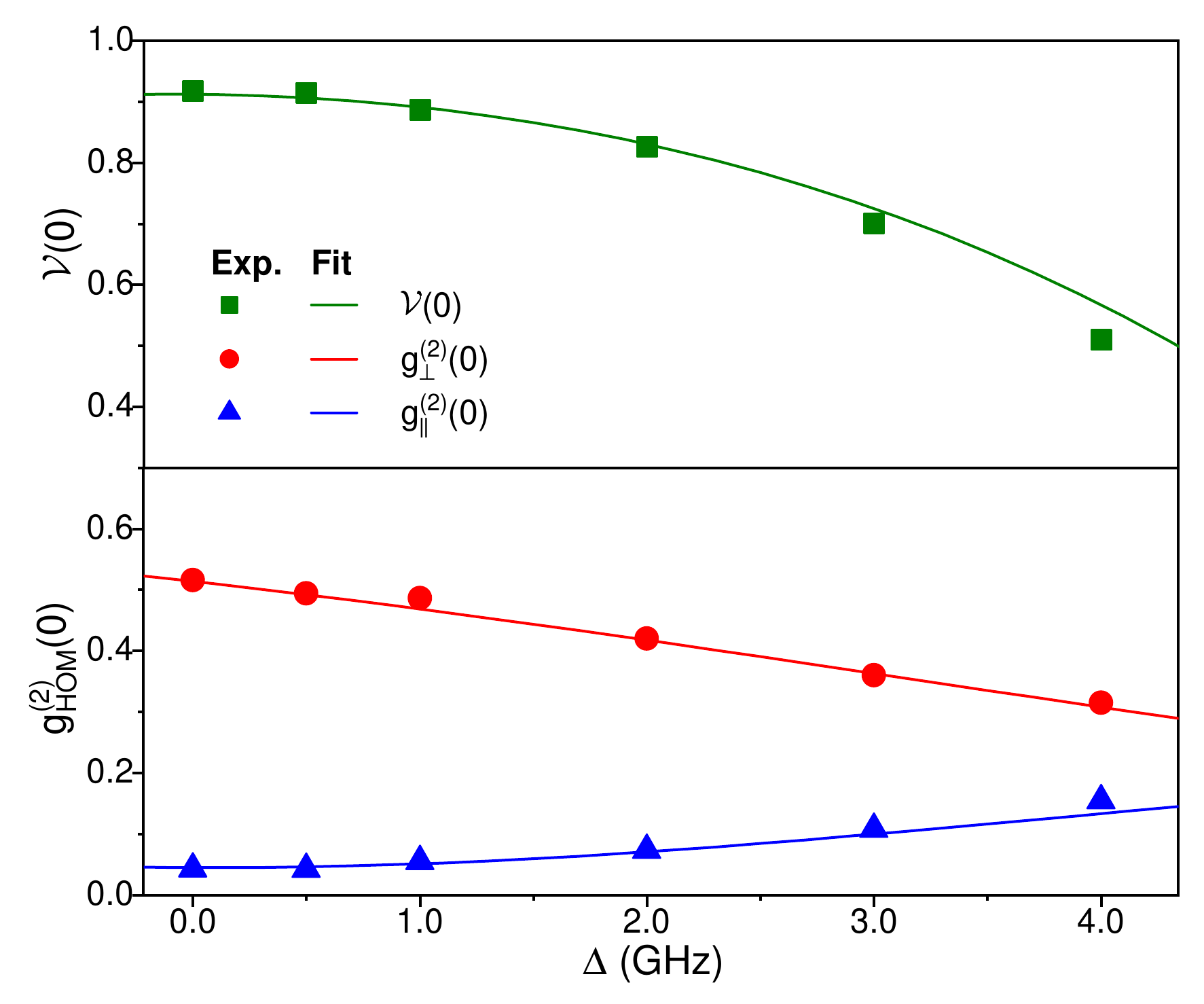}	
\caption{\textbf{Excitation detuning dependencies of the 0-delay TPI visibility and the HOM correlations.}  For all measurements, the excitation flux is fixed at $\bar{n} = 0.05$.}
	\label{fig:visibility}
\end{figure}

Fig.~\ref{fig:visibility} (bottom panel) reports the measured 0-delay HOM correlation values  
as a function of excitation detuning.  We observe a gradual increase of $g^{(2)}_{\parallel}(0)$, as expected from the RF's deteriorating single-photon purity at large excitation detunings (see Fig.~\ref{fig:setup}\textbf{b}).
In contrast, $g^{(2)}_\perp (0)$ exhibits an opposite trend, and anomalously decreases as the excitation detuning exceeds 0.5~GHz. At $\Delta =4$~GHz, $g^{(2)}_\perp (0)$ is measured to be just 0.31---well below the theoretical minimum of 0.5 for fully distinguishable photons.
Often, $g^{(2)}_\perp (0) < 0.5$ can be caused by an intensity imbalanced AMZI, which is a common flaw in a HOM  setup~\cite{Wells2023}. As an extreme example, the HOM interferometer reduces to a HBT setup when 
one AMZI arm is completely blocked. 
However, we have ruled out AMZI imbalance as the cause of the observed $g^{(2)}_\perp(0) < 0.5$.  
Instead, we attribute this anomalous behavior to a possible interplay between the large excitation detuning and the QD's fine-structure splitting. Although the exact mechanism remains to be fully understood, it likely influences both polarization configurations in a similar manner. Consequently, Eq.~\ref{eq:visibility} remains appropriate for evaluating the two-photon interference (TPI) visibility.

The experimental TPI visibility is shown as solid squares in Fig.~\ref{fig:visibility} (top panel). When $\Delta>0.5$~GHz,  $\mathcal{V}(0)$ decreases monotonically with the excitation detuning. 
The TPI visibility $\mathcal{V}(0)$
is closely linked to photon indistinguishability.  
In an ideal scenario of $g^{(2)}(0) = 0$, it is straightforward to deduce $\mathcal{V}(0) = \mathcal{M}$ from Eqs.~\ref{eq:visibility} and \ref{eq:HOM_all}. 
As all RF photons are treated as spontaneous emission, the pure-state RF model (Eq.~\ref{eq:photonstate_1}) immediately predicts their indistinguishability to be independent of excitation detuning. Apparently, the measured $\mathcal{V} (0)$ does not follow the predicted behavior for photon distinguishability. 
A number of factors can contribute to this discrepancy.   Firstly, single-photon purity degrades at large detunings due to the increasing laser background. Secondly, the detectors' time resolution prevents resolving the
HOM interference dip, especially when the dip is modulated by a large detuning frequency. Finally, the measured $g^{(2)}_\perp (0) < 0.5$ at large detunings effectively  reduce the measured $\mathcal{V}(0)$ value.
By taking into account these factors,  we construct a model (see Appendix~A3) and calculate the post-selective TPI visibility with results as the red line in Fig.~\ref{fig:visibility}. 
Here, each required $g^{(2)}(0)$ value is interpolated from the experimental measurement shown in Fig.~\ref{fig:setup}\textbf{b}.
In accordance with the RF model (Eq.~\ref{eq:photonstate_1}),  we use a fixed value of $\mathcal{M} = 0.94$, which was conservatively chosen based on the first-order coherence measurement shown in Appendix~B5.
The result (green line) is in excellent agreement with the experimental data (squares), thereby supporting the RF model based on the pure-state description.

To conclude, we have experimentally investigated post-selective TPI between mutually detuned RF signals scattered off a single QD.  By invoking the model that treats all the RF photons as a result of spontaneous emission (Eq.~\ref{eq:photonstate_1}), it is possible to find full correspondence between the predictions of the model and the experimental results.
Beyond validating the theoretical model, these findings are also highly relevant to quantum information processing, as we demonstrated that indistinguishable photons can be generated by an emitter, such as a quantum dot, irrespective of the excitation detuning.

\begin{acknowledgments}
This work was supported by the National Natural Science Foundation of China under grants 12494600, 12494604 and 62250710162, the
Beijing Natural Science Foundation under grant IS23011, and the Beijing Postdoctoral Research Foundation.

\end{acknowledgments}

\section*{Author Contributions}
\textsuperscript{\#}These authors contributed equally to this work.

\appendix

\section{Theoretical modelling}

\subsection {Model for a coherently-driven two-level system}

\begin{figure}[bt]		
\centering
\includegraphics[width=0.9\columnwidth]{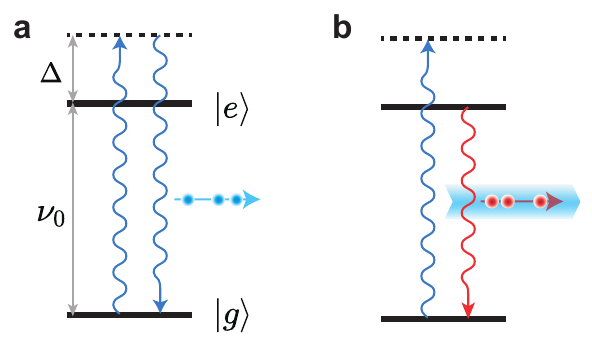}	
\caption{\textbf{Alternative pictures for a two-level emitter under weak, detuned coherent driving.} \textbf{a}, Elastic/passive scattering;
\textbf{b}, Spontaneous emission into temporal modes that are phase-locked by the excitation laser.
$\nu_0$: Transition gap between the  ground ($\ket{g}$) and excited ($\ket{e}$) states; $\Delta$: Excitation detuning.}
\label{fig:model}
\end{figure}

As depicted in Fig.~\ref{fig:model}\textbf{a}, a TLE under a weak coherent excitation is usually considered as a passive scatterer~\cite{lodahl2015}, as the scattered light inherits the driving laser's energy and linewidth (coherence).  On the other hand, the scattered signal is known to exhibit photon anti-bunching over the time scale of the TLE's radiative lifetime, which can be several orders of magnitude shorter than the driving laser's coherence time. This sharp contrast led to the term of ``sub-natural"~\cite{matthiesen2012subnatural} linewidth, and the reconciliation could be understood, with careful elaboration, in terms of the interference between the passively scattered photons and the actively emitted photons resulting from actual photon absorption and re-emission~\cite{lopezcarrenoJointSubnaturallinewidthSinglephoton2018a}.

We have recently proposed a pure-state  model~\cite{wang2023} in which all RF photons are systematically treated as an active result of absorption and re-emission.
This model successfully accounts for both the RF's first-order coherence and photon anti-bunching under strict resonant conditions.  
Extending to a detuned excitation, it immediately predicts that all RF photons from the same TLE remain identical to one another irrespective to the excitation detuning, even when the RF fields may exhibit vastly different central frequencies. Fig.~\ref{fig:model}\textbf{b} schematically summarizes the RF model under a detuned excitation: the properties of the single photons are defined by the two-level emitter but the temporal modes they radiate into are phase-locked by the excitation laser.  

For strict resonant excitation, we have previously derived the pure-state description of Eq.~\ref{eq:photonstate_1} when considering the TLE and its photon emission jointly~\cite{wang2023}, although the TLE and the emission separately are mixed states. Following similar  steps,  we now generalize this pure-state description to include the case of finite  excitation detuning.
Our starting point is the standard model~\cite{scully1997quantum} of a coherently driven two-level system, expressed in the laser-rotating frame and under continuous-wave (CW) excitation:
\begin{equation}
H=-\frac{\Omega}{2}\left(e^{-i \phi} \sigma^{\dagger}  + e^{i \phi}\sigma \right)+\frac{\Delta}{2}\sigma_z,
\end{equation}
where $\hbar=1$, $\sigma=|g\rangle \langle e|$ ($\sigma^\dagger=|e\rangle \langle g|$) is the atom lowering (raising) operator, {$\sigma_z=|g\rangle \langle g|-|e\rangle \langle e|$ is the Pauli $z$ matrix},
$\Omega$ is the Rabi frequency and $\Delta:=\nu-\nu_0$ is the excitation detuning with $\nu$ and $\nu_0$ denoting the laser and the atom transition frequencies, respectively.  
$\phi$ is the phase of the dipole matrix element $\mathbf{p}_{ge}=|\mathbf{p}_{ge}|e^{i\phi}$. 
For simplicity, and without affecting the generality of the analysis, we take $\phi = \pi/2$.
All theoretical quantities ($\Omega$, $\Delta$, and $\gamma$) are expressed in angular frequency units (rad s$^{-1}$), whereas the corresponding parameters in the Main Text are quoted as cyclic frequencies (Hz).

If we consider the spontaneous emission with the Markov approximation, we have
\begin{equation}
\frac{d}{d t} \rho=\mathcal{L} \rho=-i[H, \rho]+{\color{black} \mathcal{D}_{\gamma_{\parallel},\sigma}[\rho_a]+ \mathcal{D}_{\gamma^*/2,\sigma_z}[\rho_a]},
\end{equation}
with $\gamma_\|$ being the spontaneous emission rate, $\gamma^*$ the pure dephasing rate of the emitter, $\rho_a$ is the density matrix of a two-level atom, i.e., TLE and $\mathcal{D}$ the dissipation superoperator given by
\begin{equation}
\mathcal{D}_{x,A}[\rho_a] =x \left[A \rho A^{\dagger}-\frac{1}{2} (A^{\dagger} A \rho+ \rho A^{\dagger} A)\right].
\end{equation}

From an initial state $\rho(0)=|g\rangle\langle g|$, we analytically solve the system state at time $t$ by

\begin{equation}
\rho(t)=e^{t \mathcal{L}} \rho(0).
\end{equation}

Since we are only interested in the system at equilibrium,  we can simplify our analysis by considering the steady-state value as $t\rightarrow \infty$ to get
\begin{eqnarray}
\rho_{g g}(\infty)=zy{\frac{2\gamma_{\parallel}\left(\gamma_\perp^2+ \Delta^2\right)+\gamma_{\perp}\Omega^2}{2\gamma_{\parallel}\left(\gamma_\perp^2+ \Delta^2\right)+2\gamma_{\perp}\Omega^2}}, \\
\rho_{e e}(\infty)=\frac{\gamma_\perp \Omega^2}{2\gamma_{\parallel}\left(\gamma_\perp^2+ \Delta^2\right)+2\gamma_\perp \Omega^2}, \\
\rho_{g e}(\infty)=\rho_{e g}^*(\infty)=\frac{\gamma_{\parallel}\left(\gamma_\perp+ i \Delta\right) \Omega}{2\gamma_{\parallel}\left(\gamma_\perp^2+ \Delta^2\right)+2\gamma_\perp \Omega^2},
\end{eqnarray}
\noindent where $\gamma_\perp=\gamma_{\parallel}/2+\gamma^*$ is the spectrally-broadened emitter linewidth. Then, the steady-state density matrix of the TLE alone can be written as:
\begin{widetext}
\begin{equation}
\begin{aligned}
\hat{\rho}_a(t \rightarrow \infty)=\frac{1}{2 \gamma_{\|}\left(\gamma_{\perp}^2+\Delta^2\right)+2 \gamma_{\perp} \Omega^2}\{ & {\left[2 \gamma_{\|}\left(\gamma_{\perp}^2+ \Delta^2\right)+\gamma_{\perp} \Omega^2\right]|g\rangle\langle g|+\gamma_{\|}\left(\gamma_{\perp}+i \Delta\right) \Omega|g\rangle\langle e| } \\
& \left.+\gamma_{\|}\left(\gamma_{\perp}-i \Delta\right) \Omega|e\rangle\langle g|+\gamma_{\perp} \Omega^2|e\rangle\langle e|\right\}.
\end{aligned}
\end{equation}
\end{widetext}

With the hypothesis that the driving laser is perfectly filtered from the radiation emitted by the source, we can leverage the relation of proportionality between the source dipole and the field operator of the collected light, i.e., $a \propto \sigma$. Therefore, we might as well let $a=B\sigma$, where $B$ is a proportional coefficient.
In this way, we always have that the state of light normalized by the intensity corresponds to the steady-state of the emitter.

Formally, we can perform a tomographic light analysis by examining its coherence and intensity at a specific instant $t$, once the emitter has reached a steady state:
\begin{equation}
\langle a(t)\rangle_{\infty}=\operatorname{Tr}\left[B\sigma \rho_{\infty}\right]=B\rho_{eg}(\infty)
\end{equation}
\begin{equation}
\left\langle a^{\dagger}(t) a(t)\right\rangle_{\infty}=\operatorname{Tr}\left[B^2 \sigma^{\dagger} \sigma \rho_{\infty}\right]=B^2 \rho_{ee}(\infty).
\end{equation}
The first order coherence is then obtained by normalizing the squared magnitude of the coherence by the intensity:
\begin{equation}
|g_{\infty}^{(1)}|=\frac{|B\rho_{eg}(\infty)|^2}{B^2\rho_{ee}(\infty)}=\frac{\gamma_\parallel^2\left(\gamma_\perp^2+ \Delta^2\right)}{2\gamma_\perp\left[\gamma_\parallel\left(\gamma_\perp^2+\Delta^2\right)+\gamma_\perp\Omega^2\right]}.
\end{equation}
It can be seen that the specific value of $B$ does not affect $g^{(1)}_\infty$ of fluorescence. Since a single excitation of the TLE can be converted into a single photon, thus the fluorescence density matrix at the steady state can be determined as well,
\begin{widetext}
\begin{equation}
\begin{aligned}
\hat{\rho}_p(t \rightarrow \infty)=\frac{1}{2 \gamma_{\|}\left(\gamma_{\perp}^2+\Delta^2\right)+2 \gamma_{\perp} \Omega^2}\{ & {\left[2 \gamma_{\|}\left(\gamma_{\perp}^2+\Delta^2\right)+\gamma_{\perp} \Omega^2\right]|0\rangle\langle 0|+\gamma_{\|}\left(\gamma_{\perp}+i \Delta\right) \Omega|0\rangle\langle 1| } \\
& \left.+\gamma_{\|}\left(\gamma_{\perp}-i \Delta\right) \Omega|1\rangle\langle 0|+\gamma_{\perp} \Omega^2|1\rangle\langle 1|\right\}.
\end{aligned}
\end{equation}
\end{widetext}

After obtaining the reduced density matrices of the TLE and fluorescence individually, we can infer the density matrix of the TLE-photon joint system.
Under the equilibrium condition of continuous-wave laser driving, at any given time $t$, the density matrix state vector of the joint system of the two-level atom and fluorescence only needs to consider $|g0\rangle$, $|e0\rangle$, and $|g1\rangle$, while excluding higher-order terms such as $|e1\rangle$ and $|g2\rangle$. This is primarily based on exciting experimental observations and theoretical results, where $g^{(2)}(0)=0$ is always valid, regardless of whether the excitation is strong or weak. 
In the absence of pure dephasing, i.e., $\gamma_\perp=\gamma_\|/2$, 
the instantaneous atom-photon density matrix can be inferred to have 
the form

\begin{widetext}
\begin{equation}
\hat{\rho}_{ap}=\left(\begin{array}{ccc}
\rho_{g 0, g 0} & \rho_{g 0, g 1} & \rho_{g 0, e 0} \\
\rho_{g 1, g 0} & \rho_{g 1, g 1} & \rho_{g 1, e 0} \\
\rho_{e 0, g 0} & \rho_{e 0, g 1} & \rho_{e 0, e 0}
\end{array}\right)
=\frac{1}{\gamma_{\|}^2+4\Delta^2+2 \Omega^2}\left(\begin{array}{ccc}
\gamma_{\parallel}^{2} + 4\Delta^{2} & (\gamma_{\parallel} + 2i\Delta)\Omega & (\gamma_{\parallel} + 2i\Delta)\Omega \\
(\gamma_{\parallel} -2i\Delta)\Omega & \Omega^{2} & \Omega^{2} \\
(\gamma_{\parallel} -2i\Delta)\Omega & \Omega^{2} & \Omega^{2}\end{array}\right).
\end{equation}
\end{widetext}
The specific structure of $\hat{\rho}_{ap}$—with all nonzero elements phase-coherent and rank one—reveals that the joint atom–photon system remains in a pure quantum state.
Consequently, the above density matrix can be equivalently expressed in the pure-state form
\begin{equation}
\ket{\psi}
= \sqrt{p_0}\ket{0,g} + \sqrt{p_1} e ^ {-i \theta^\prime} \frac{\ket{0,e} + \ket{1,g}}{\sqrt{2}},
\end{equation}
where $\theta^\prime = \arctan\left(\frac{2\Delta}{\gamma_{\|}}\right)$ and
\begin{align}
p_0 &= \frac{\gamma_{\parallel}^2 + 4\Delta^2}{\gamma_{\parallel}^2 + 4\Delta^2 + 2\Omega^2}, \
p_1 = \frac{2\Omega^2}{\gamma_{\parallel}^2 + 4\Delta^2 + 2\Omega^2}.
\end{align}

By transforming back from the laser-rotating frame to the laboratory frame, we obtain the exact form of Eq.~\ref{eq:photonstate_1} for the joint TLE–photon system,  
\begin{equation}
 \ket{\psi}
 = \sqrt{p_0}\ket{0,g}
 + \sqrt{p_1}\,e^{-i(2\pi\nu t+\theta)}\,
   \frac{\ket{0,e}+\ket{1,g}}{\sqrt{2}},
 \label{eq:photonstate_2}
\end{equation}
where $\theta = \arctan\left(\frac{2\Delta}{\gamma_{\parallel}}\right) + \frac{\pi}{2}$, reinstating the optical oscillation at frequency~$\nu$ and the dipole phase initially set at $\pi/2$.

\subsection{Coefficients $p_0$ and $p_1$}

The dependencies of $p_0$ and $p_1$ on the excitation power~$\bar{n}$ can be derived from the balance between photon absorption and re-emission by the TLE~\cite{wang2023}, yielding
\begin{equation}
    p_0 = \frac{1}{1 + 2\eta_{ab}\bar{n}},  \qquad
    p_1 = \frac{2\eta_{ab}\bar{n}}{1 + 2\eta_{ab}\bar{n}},
\label{eq:p0_appendix}
\end{equation}
where $\eta_{ab}$ denotes the TLE’s absorption efficiency in the weak-drive regime, which depends on the excitation detuning.  
In the weak-drive limit, $p_0 \!\to\! 1$, indicating that the emitter predominantly remains in its ground state.  

The connection between the coefficients $p_0$ and $p_1$ and the RF power spectrum can be established by evaluating the first-order coherence.  
Starting from the pure state in Eq.~\ref{eq:photonstate_2}, the normalized first-order coherence function is
\begin{equation}
g^{(1)}(\tau)
=\frac{\langle \psi|a^{\dagger}(t)a(t+\tau)|\psi\rangle}
{\langle \psi|a^{\dagger}a|\psi\rangle}.
\label{eq:g1_def}
\end{equation}
Evaluating this expression gives
\begin{equation}
g^{(1)}(\tau)=p_0\,e^{-i2\pi\nu\tau},
\label{eq:g1_result}
\end{equation}
valid for $T_1\!\ll\!\tau\!\ll\!T_L$.  
Since the RF power spectrum is the Fourier transform of $g^{(1)}(\tau)$, the phase-coherent term $p_0\,e^{-i2\pi\nu\tau}$ corresponds to a narrow, laser-like spectral component centered exactly at the excitation frequency~$\nu$ with a relative weight~$p_0$, whereas the remaining fraction~$p_1$ gives rise to the broadband background.  

The laser-like and broadband fractions in the RF spectrum are conventionally referred to as the coherent and incoherent scattering components, respectively~\cite{Steck2023}.  
Their normalized spectral weights are usually expressed as
\begin{equation}
    I_{\mathrm{coh}} = \frac{1}{1+s}, \qquad
    I_{\mathrm{incoh}} = \frac{s}{1+s},
\label{eq:s-parameter}
\end{equation}
where $s = \frac{2\Omega^2 / \gamma_\|^2}{1 + (2\Delta / \gamma_\|)^2}$ is the saturation parameter.  
By identifying $s = 2\eta_{ab}\bar{n}$, Eq.~\ref{eq:s-parameter} becomes identical to Eq.~\ref{eq:p0_appendix}, leading to $p_0 = I_{\mathrm{coh}}$ and $p_1 = I_{\mathrm{incoh}}$. This equivalence establishes a one-to-one correspondence between the conventional mixed-state RF description and the pure-state formulation we have introduced, where the joint emitter--photon system remains in a coherent superposition of the vacuum and one-photon states.


\subsection{$g^{(2)}_\parallel (\tau)$ under steady-state condition ($\tau \gg T_1$)}

In this section, we will discuss the coincidence counts for different incident configurations depicted in Fig.~\ref{fig:TPI-Schematic} from the perspective of the input photon states.
We start from the pure-state description of  Eq.~\ref{eq:photonstate_1}.  Tracing out the TLE component states ($\ket{g}$ and $\ket{e}$), and neglecting the detuning-dependent phase $\theta$, the remaining photon subsystem is described by a mixed state. 
This mixed state is equivalent to the photon-number coherent state
\begin{equation}
\ket{\psi}_{\mathrm{ph}} = \sqrt{p_0}\ket{0} + \sqrt{p_1}\,e^{-i2\pi\nu t}\ket{1},
\end{equation}
attenuated by an effective channel with 50\% loss.  Conveniently, channel losses do not affect the analysis of the first-order or second-order correlation functions as we have previously proved~\cite{wang2023}, so we are able to use the pure photon state in the subsequent derivation.

We consider two mutually-detuned RF signals entering a 50/50 beam splitter via ports $a$ and $b$, and exiting from ports $c$ and $d$. Under steady-state condition ($\tau \gg T_1$), the input state can be expressed as the product of photon states at different instants of time for ports $a$ and $b$ as $|\psi_{in}\rangle=|\psi\rangle_{a}|\psi\rangle_{b}$, where:
\begin{widetext}
\begin{equation}
\begin{aligned}
&|\psi\rangle_{a}=|\psi\rangle_{a,t}|\psi\rangle_{a,t+\tau}=\left(\sqrt{p_{0}}|0\rangle_{a,t}+\sqrt{p_{1}}|1\rangle_{a,t}e^{-i2\pi\nu_{a}t}\right)\left(\sqrt{p_{0}}|0\rangle_{a,t+\tau}+\sqrt{p_{1}}|1\rangle_{a,t+\tau}e^{-i2\pi\nu_{a}(t+\tau)}\right),
\\&|\psi\rangle_{b}=|\psi\rangle_{b,t}|\psi\rangle_{b,t+\tau}=\left(\sqrt{p_{0}}|0\rangle_{b,t}+\sqrt{p_{1}}|1\rangle_{b,t}e^{-i2\pi\nu_{b}t}\right)\left(\sqrt{p_{0}}|0\rangle_{b,t+\tau}+\sqrt{p_{1}}|1\rangle_{b,t+\tau}e^{-i2\pi\nu_{b}(t+\tau)}\right).
\end{aligned}
\end{equation}
In the above formula, $\nu_{a}=\nu_0-\frac{\Delta}{2};$ $\nu_{b}=\nu_0+\frac{\Delta}{2}$, $\Delta$ is the frequency detuning relative to the center frequency $\nu_0$ and $\tau$ is the coincidence time interval.
According to the transformation relation of a beam splitter, the input optical field annihilation operator can be given by
\begin{equation}
\begin{aligned}
    c_td_{t+\tau}&=\frac{1}{\sqrt2}(a_t+b_t)\times\frac{1}{\sqrt2}(a_{t+\tau}-b_{t+\tau})
    \\&=\frac{1}{2}(a_ta_{t+\tau}-a_tb_{t+\tau}+a_{t+\tau}b_t-b_tb_{t+\tau}),
\end{aligned}
\end{equation}
where $a_t$ and $a_{t+\tau}$ act on states $|\psi\rangle_{a,t}$ and $|\psi\rangle_{a,t+\tau}$. Similarly, the remaining annihilation operators adhere to this convention. Then, the count probabilities of detectors in ports $c$ and $d$ can be written as
\begin{equation}
\begin{aligned}
    P(\tau)&=\langle\psi_{in}|c_t^\dagger d_{t+\tau}^\dagger
    c_td_{t+\tau}|\psi_{in}\rangle\\
 & =\left|\frac{1}{2}a_ta_{t+\tau}|\psi_{in}\rangle\right|^2+\left|-\frac{1}{2}b_tb_{t+\tau}|\psi_{in}\rangle\right|^2+\left|-\frac{1}{2}a_tb_{t+\tau}|\psi_{in}\rangle+\frac{1}{2}a_{t+\tau}b_t|\psi_{in}\rangle\right|^2 \\
 & =\left|\frac{1}{2}p_1e^{-i2\pi \nu_a(2t+\tau)}|00\rangle_a|\psi\rangle_b\right|^2+\left|-\frac{1}{2}p_1e^{-i2\pi \nu_b(2t+\tau)}|00\rangle_b|\psi\rangle_a\right|^2\\
 & +\left|-\frac{1}{2}(p_0p_1e^{-i2\pi2\nu t}e^{-i2\pi2\nu_b\tau}|0000\rangle+ p_1\sqrt{p_0p_1}e^{-i2\pi2\nu t}e^{-i2\pi2\nu_b(t+\tau)}|0010\rangle +p_1\sqrt{p_0p_1}e^{-i2\pi2\nu(t+\tau)}e^{-i2\pi2\nu_at}|0100\rangle\right.\\& \left.+p_1{}^2e^{-i2\pi2\nu(2t+\tau)}|0110\rangle)+\frac{1}{2}(p_0p_1e^{-i2\pi2\nu t}e^{-i2\pi2\nu_a\tau}|0000\rangle+p_1\sqrt{p_0p_1}e^{-i2\pi2\nu(t+\tau)}e^{-i2\pi2\nu_b\tau}|0001\rangle\right.\\& \left.+p_1\sqrt{p_0p_1}e^{-i2\pi2\nu t}e^{-i2\pi2\nu_a(t+\tau)}|1000\rangle+{p_1}^2e^{-i2\pi2\nu(2t+\tau)}|1001\rangle)\right|^2\\
 & =\frac{1}{4}p_{1}^{2}+\frac{1}{4}p_{1}^{2}+\frac{1}{2}p_{1}^{2}\left[1-p_{0}^{2}\cos{(2\pi\Delta\tau)}\right],
\end{aligned}
\end{equation}
\end{widetext}
where the form of photon states (e.g., $\ket{0000}$) follows the convention that the first two digits correspond to the $t$ and $t+\tau$ time instants of port $a$ respectively, and the last two digits correspond to the $t$ and $t + \tau$ instants of port $b$. Within the final equation, the first and second terms correspond to cases (\textbf{i}) and (\textbf{iv}) illustrated in Fig.~\ref{fig:TPI-Schematic}, while the last term accounts for the interference between the two indistinguishable paths as shown in Fig.~\ref{fig:TPI-Schematic} cases (\textbf{ii}) and (\textbf{iii}). Considering finite photon indistinguishability, we obtain the second-order HOM correlation function by normalizing to the coincidence probability of $p_1^2$ at $\tau = \infty$:
\begin{equation}
    g_{HOM}^{(2)}\left( \Delta \right) = 1 - \frac{1}{2}\mathcal{M}p_{0}^{2}\cos2\pi\Delta\tau,~| \tau| \gg T_1,
\end{equation}
which is identical to Eq.~\ref{eq:HOM_Async} in Main Text.

\subsection{Modeling the 0-delay HOM results}

Here, we describe the calculation method that we use to fit the experimental data of the excitation detuning dependence of the 0-delay TPI visibility $\mathcal{V}(0)$ shown in Fig.~\ref{fig:visibility}.

Each experimental $\mathcal{V}(0)$ value is calculated from the corresponding HOM measurements of 
$g^{(2)}_{\perp}(0)$ and $g^{(2)}_{\parallel}(0)$ using to Eq.~\ref{eq:visibility}. However,  the measured $g^{(2)}_\perp (0)$ values at large detunings fall below the theoretical minimum of 0.5 for completely distinguishable but intensity-balanced RF inputs.
To account for this irregularity ($g^{(2)}_\perp < 0.5$), we introduce an empirical parameter $x$ ($0 \leq x \leq 1$). We fit the measured HOM correlation data using
\begin{equation}
    g^{(2)}_{HOM} (\tau)  = \left [ x \times g^{(2)} (\tau) + (1-x)  \times g\prime^{(2)}_{HOM}(\tau) \right ]  \otimes R(\tau),
\end{equation}
where $g\prime^{(2)}_{HOM}$ represents the HOM correlation function as described by  Eq.~\ref{eq:HOM_all} for always balanced RF signals and the inequality $g\prime^{(2)}_{\perp} (0) \geq 0.5$ holds. $R(\tau)$ represents the detector time response. 

Among simulation parameters, $x$ is a fitting parameter and excitation detuning dependent.  The auto-correlation functions $g^{(2)}(\tau)$ used are obtained from the deconvolution of the measured HBT data with the detector response $R(\tau)$. We use fixed values of $\mathcal{M}_\perp = 0$ and $\mathcal{M}_\parallel = 0.94$ for all excitation fluxes. $R(\tau)$ is a Gaussian function of  59~ps FWHM.

\section{Further experimental details}

\subsection{Pulse-carving the excitation lasers}

\begin{figure}[hbp]		
\centering
\includegraphics[width=0.8\columnwidth]{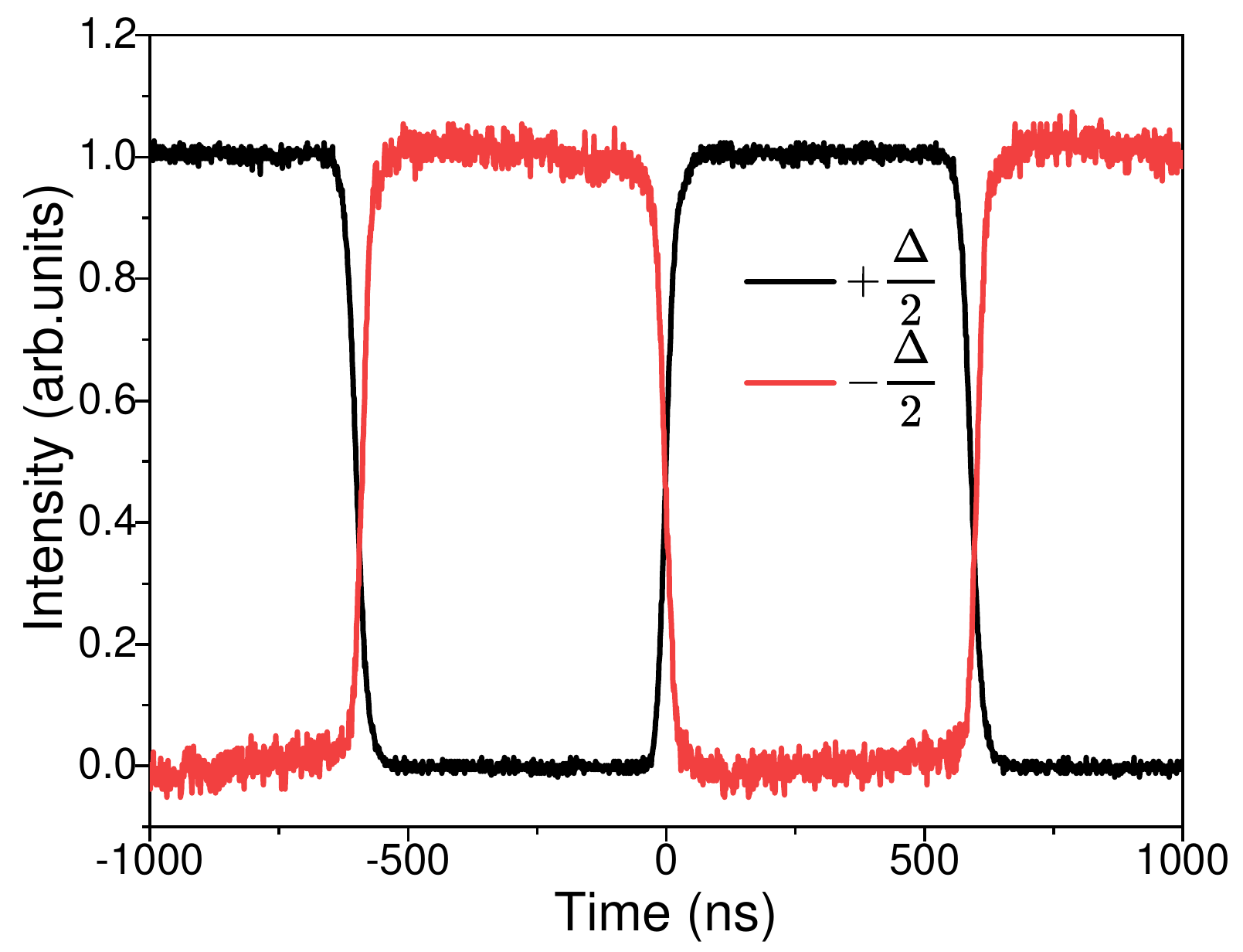}	
\caption{\textbf{Red and blue detuned excitation laser waveforms.}}
\label{fig:fig10}
\end{figure}

Two continuous-wave lasers are modulated into square-wave signals with a period of approximately 1190 ns using two independent acousto-optic modulators (AOMs), each providing an extinction ratio of about 50 dB. The AOMs are driven by synchronized trigger signals from the same function generator, with a relative phase shift of half a period between them. The chopped square pulses are detected by a photodetector and recorded on an oscilloscope, as shown in Fig.~\ref{fig:fig10}. It is observed that the two lasers alternate with a switching period of 595\,ns. The electrical 20--80\% edge times are measured as $\tau_{\rm r} \approx $24 ns (rise) and $\tau_{\rm f}\approx$30 ns (fall), respectively.   
These values satisfy $\tau_{\rm r},\tau_{\rm f}\gg T_1$ and $1/\tau_{\rm r,f}\ll \Delta_{\max}=4\,$GHz, suppressing edge-induced spectral artifacts while keeping the cross-talk between adjacent tones below $\text{1--2}\%$. 
In addition, the energy detuning between the two square waves is determined by measuring the beat frequency between the two lasers using a frequency counter.

\subsection{Radiative lifetime $T_1$ under cavity resonance}
\begin{figure}[hbp]		
\centering
\includegraphics[width=0.8\columnwidth]{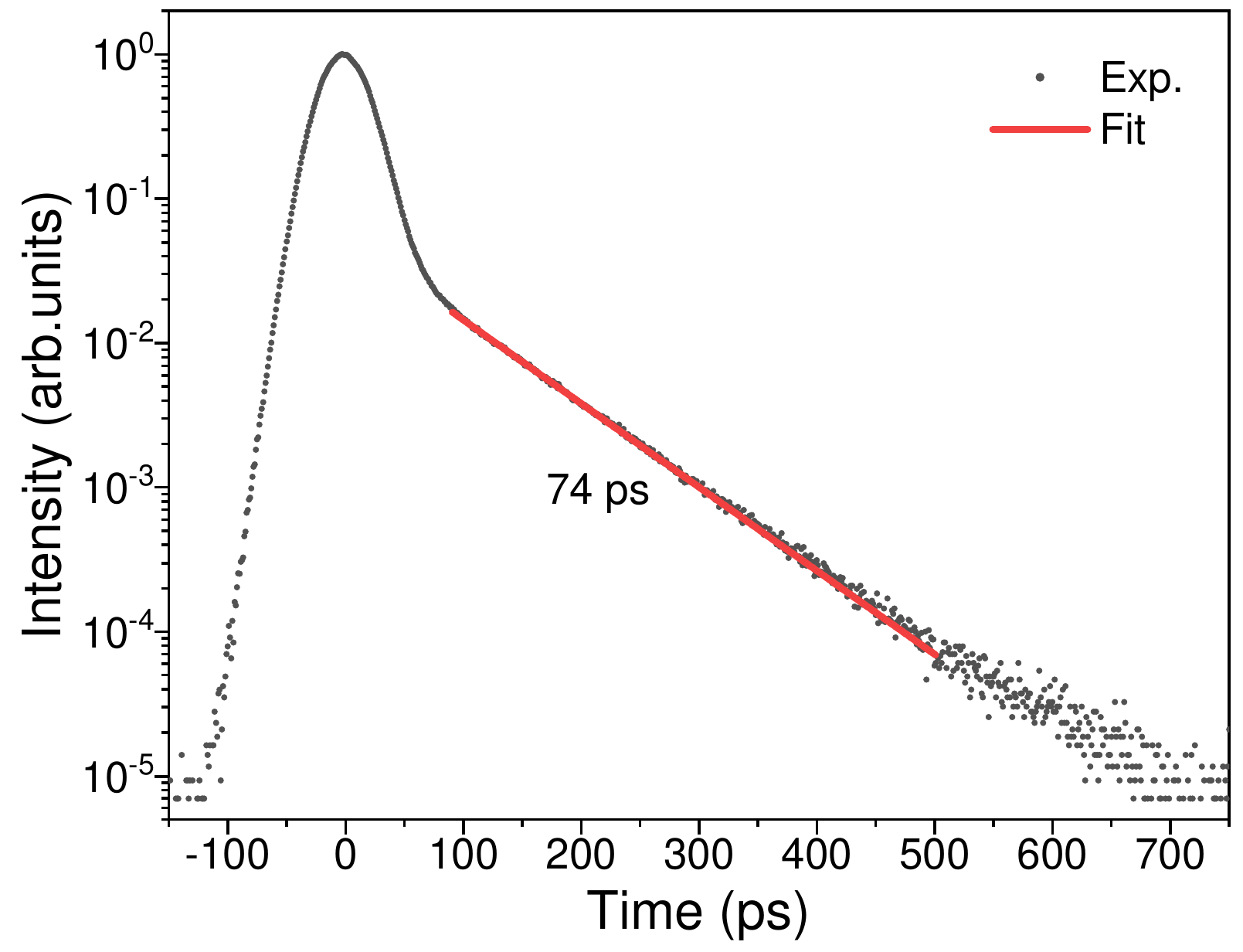}	
\caption{\textbf{Time-resolved resonance fluorescence measurement when the QD is tuned into resonance with the cavity.} The black dots represent the experimental data, and the green line shows a single-exponential fit, yielding a radiative lifetime of $T_1 = 74$~ps.}
\label{fig:fig9}
\end{figure}

The radiative lifetime $T_1$ is a key parameter characterizing the device performance. In this study, we use $T_1$ directly to calculate the excitation flux $\bar{n}$. To measure $T_1$, we employ a time-resolved setup  using an 80~MHz picosecond mode-locked laser to resonantly excite the QD~\cite{wu2023}. The measured data are presented in Fig.~\ref{fig:fig9}. The time-resolved spectrum exhibits a pronounced laser background around zero delay, followed by an exponential decay component attributed to the QD emission. By fitting this decay, we extract a radiative lifetime of $T_1 = 74$~ps. In this measurement, the QD is tuned into resonance with the microcavity.

\subsection{Fine-structure splitting (FSS)}

\begin{figure}[hbp]		
\centering
\includegraphics[width=.9\columnwidth]{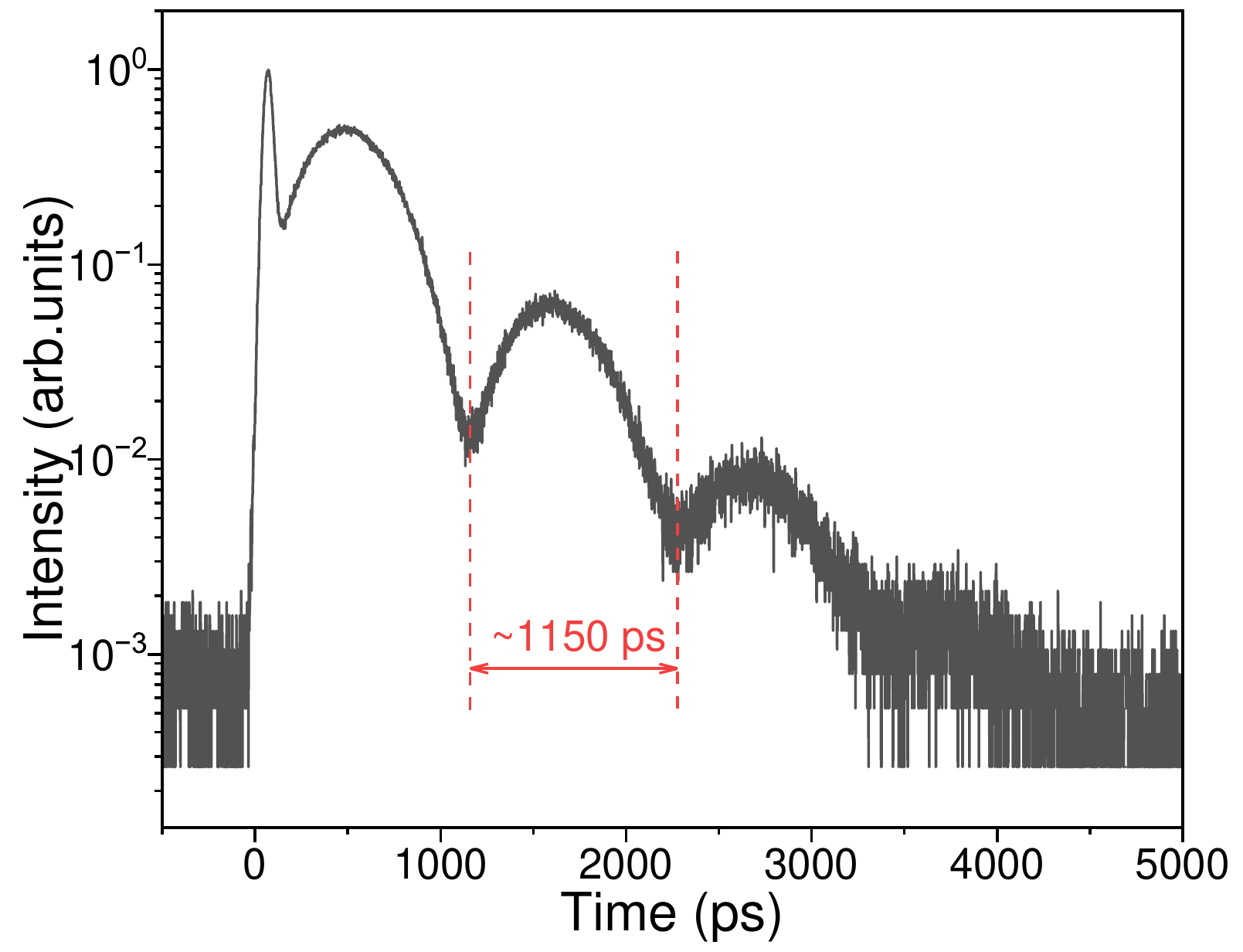}	
\caption{\textbf{Time-resolved PL spectrum of the QD detuned from the cavity mode.} The red dashed line highlights the oscillation period.}
\label{fig:fig11}
\end{figure}

\begin{figure*}[hbt]		
\centering
\includegraphics[width=1.8\columnwidth]{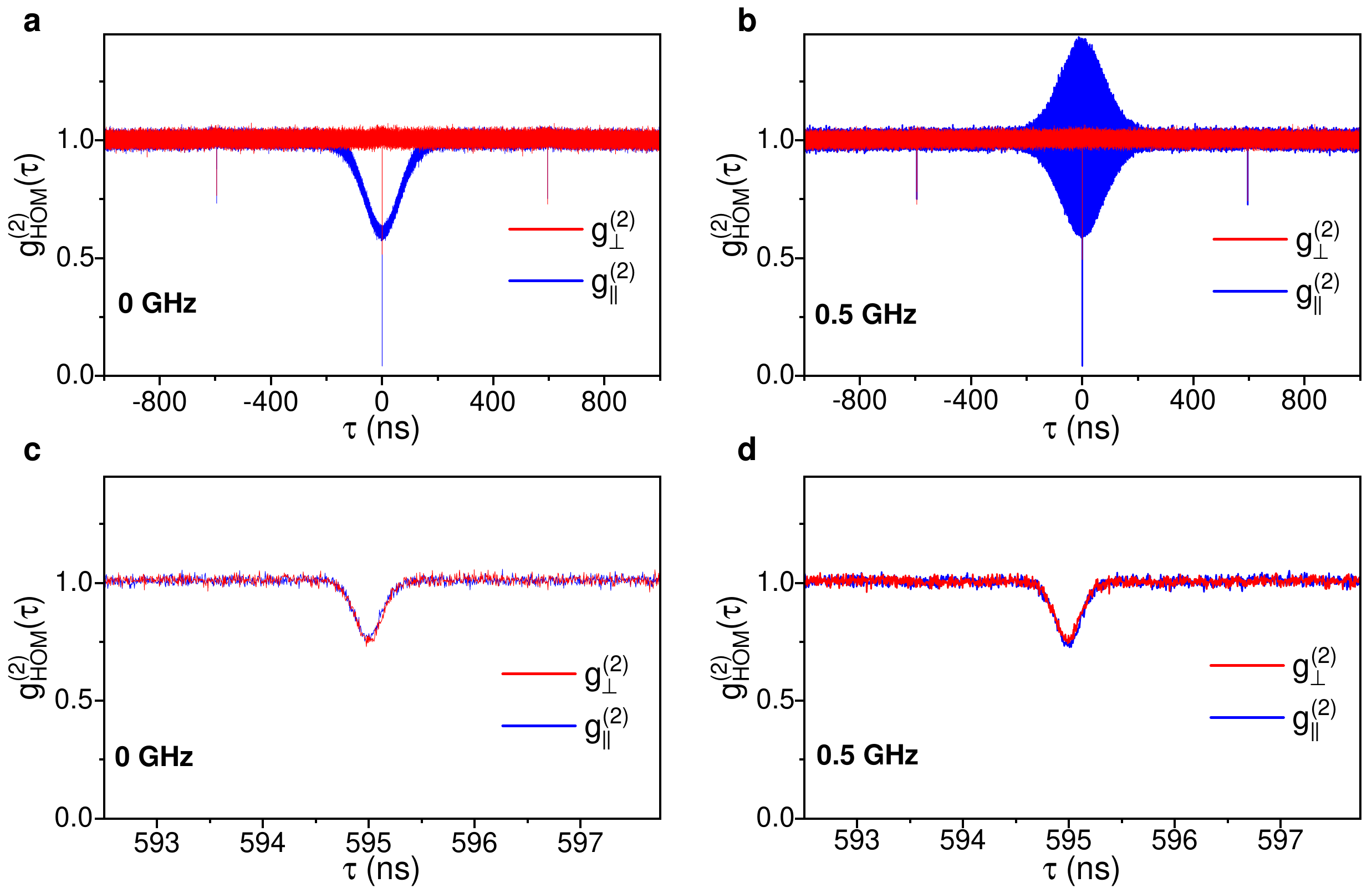}	
\caption{\textbf{Dual-color HOM interference traces over an extended delay range under two different mutual detunings: $\bm{\Delta =0}$~GHz (left column) and $\bm{\Delta =0.5}$~GHz (right column).}
\textbf{a}, \textbf{b}, HOM interference traces for the parallel (blue) and orthogonal (red) polarization settings;  \textbf{c}, \textbf{d}, Magnified view of the data shown in panels \textbf{a} and \textbf{b}.}
\label{fig:fig8}
\end{figure*}

The neutral exciton nature of the QD transition is confirmed by the presence of fine-structure splitting (FSS)~\cite{gammon1996}. 
To measure the splitting, we employ cross-polarized resonant excitation, with the excitation polarization chosen such that both exciton eigenstates are equally addressed. 
Due to the FSS, the temporal evolution of the collected signal exhibits oscillations arising from the energy difference between the two polarsation states~\cite{ollivier2020reproducibility}. 
By extracting the oscillation period, we determine the fine-structure splitting $\Delta_\text{FSS}$. 
To better resolve the oscillations, we detune the QD transition off the cavity resonance by raising the device temperature. 
The resulting temporal dynamics are presented in Fig.~\ref{fig:fig11}, where the oscillation period, highlighted by the red dashed line, is $\sim 1150$~ps, corresponding to $\Delta_\text{FSS} \approx 0.91$~GHz.

\subsection{HOM results at the AMZI delay}

In Fig.~\ref{HOM} of the main text, we plotted the dual-color HOM interference over a delay range of $\pm 400$~ns. To ensure completeness, Fig.~\ref{fig:fig8} presents the same dataset over an extended delay range, revealing the correlation features at the characteristic delays of $\pm 595$~ns. A zoomed-in view around $+595$~ns is also included for clarity.

Here, the 595~ns delay significantly exceeds the mutual coherence time of the excitation laser sources ($T_c \approx 200$~ns), thereby erasing any phase correlation between the photon pairs. Consequently, no interference effect is observable at this delay. As shown in Figs.~\ref{fig:fig8}c and \ref{fig:fig8}d, regardless of whether the excitation laser is detuned or the polarization is matched, the observed correlation converges to a dip value of 0.75. This quantitatively matches the results observed in single-photon streams under incoherent excitation conditions~\cite{patelPostselectiveTwoPhotonInterference2008}.

It is also worth noting that the delay configuration of the AMZI used here is fundamentally different from that in our prior work~\cite{wang2023}, where the AMZI delay (4.92~ns) was much shorter than the coherence time of the excitation laser. In that case, interference effects manifested prominently at the side peaks of the HOM interference traces. These rich experimental features were discussed in detail in Ref.~\cite{wang2023}.

\subsection{Quantifying the attainable photon indistinguishability}

\begin{figure}[hbt]		
\centering
\includegraphics[width=0.75\columnwidth]{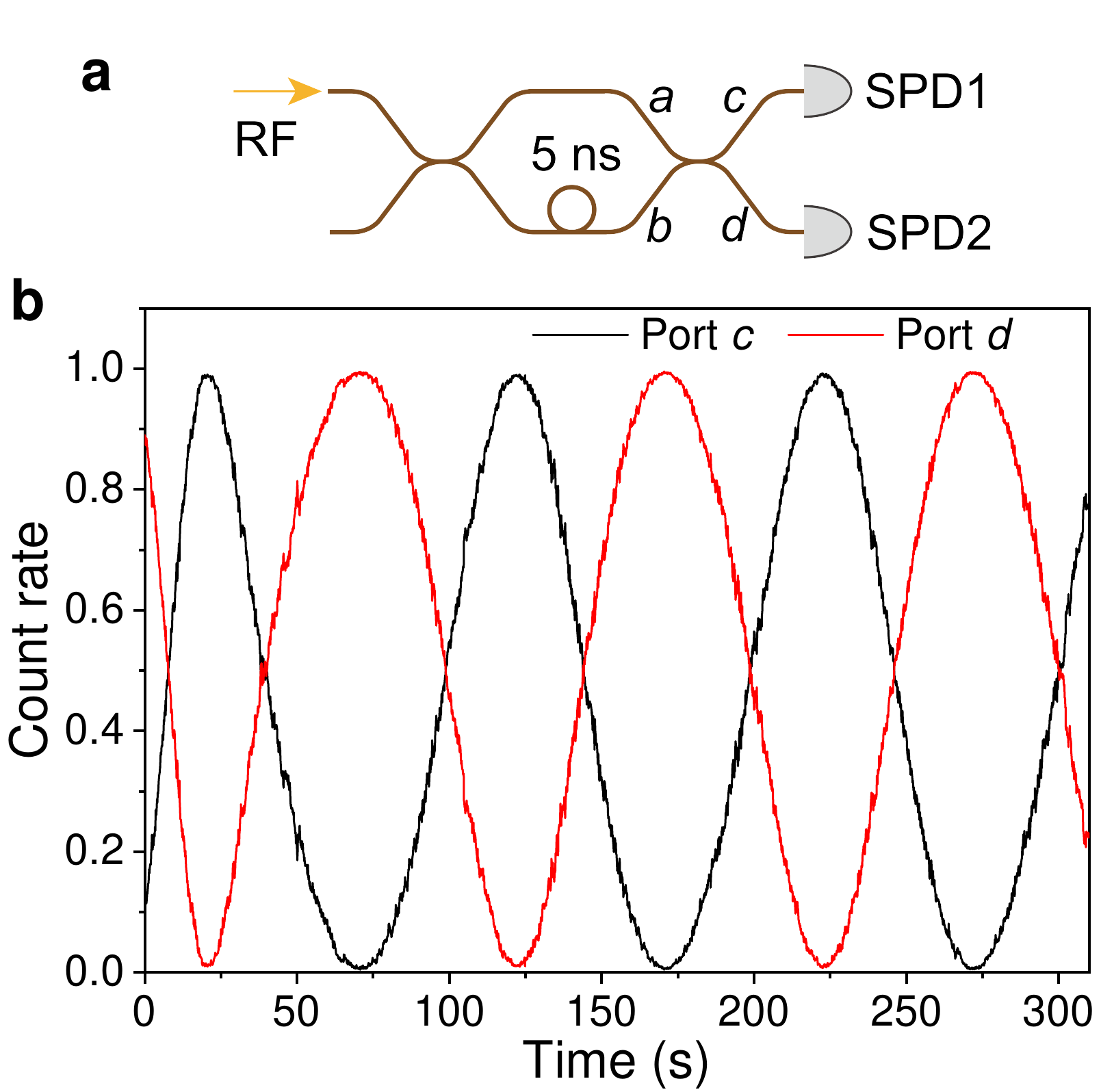}	
\caption{\textbf{The RF's first-order interference.} \textbf{a}, Purpose-built AMZI with a differential delay of 5~ns; \textbf{b}, Normalized count rates at ports $c$ (black) and $d$ (red) under a single laser excitation of $\Delta$ = 0~GHz and $\bar{n}$ = 0.002. The phase drift is caused by varying the temperature of the AMZI.}
\label{fig:g1}
\end{figure}

Photon indistinguishability $\mathcal{M}$ is an important parameter in simulating the HOM correlation traces shown in Fig.~\ref{HOM} as well as the detuning-dependent TPI visibility in Fig.~\ref{fig:visibility}. We have chosen a fixed value of $\mathcal{M} = 0.94$, and below we justify this choice through measuring the RF's first-order correlation function $g^{(1)}$.

In measuring $g^{(1)}$, we use a second, purpose-built AMZI with a differential delay of $\tau =5$~ns as shown in Fig.~\ref{fig:g1}\textbf{a}. This delay is much greater than the QD relaxation time but much shorter than the laser coherence time.  Under this condition of $T_1 \ll \tau \ll T_L$, the RF's first-order correlation function can be written as~\cite{de_santis_solid-state_2017,wang2023}
\begin{equation}
|g^{(1)}(\tau)| = \sqrt{\mathcal{M}}p_0,
\end{equation}
from which we can bound photon indistinguishability to 
$\mathcal{M} \leq |g^{(1)}|^2 \times (1 + 2 \bar{n})^2$ using Eq.~\ref{eq:p0} and  inequality $\eta_{ab} \leq 1$. 

Fig.~\ref{fig:g1}\textbf{b}
shows the interference fringes, measured with an excitation flux of $\bar{n} = 0.002$ under strictly resonant excitation. 
We extract a first-order correlation value of $|g^{(1)}| = 0.991 \pm 0.008$, which sets an upper bound of $\mathcal{M} \leq 0.990$ for the attainable indistinguishability.

%

\end{document}